\documentclass[twocolumn,twocolappendix]{aastex631}

\newcommand{\jhrev}[1]{{\color{black}#1}}

\shorttitle{Shadowing of ZZ Tau IRS}
\shortauthors{Hashimoto et al.}

\graphicspath{{./}{figures/}}

\begin{document}

\title{Shadowing in the protoplanetary disk of ZZ Tau IRS with HST}

\correspondingauthor{Jun Hashimoto}
\email{jun.hashimto@nao.ac.jp}

\author[0000-0002-3053-3575]{Jun Hashimoto}
\affil{Astrobiology Center, National Institutes of Natural Sciences, 2-21-1 Osawa, Mitaka, Tokyo 181-8588, Japan}
\affil{Subaru Telescope, National Astronomical Observatory of Japan, Mitaka, Tokyo 181-8588, Japan}
\affil{Department of Astronomy, School of Science, Graduate University for Advanced Studies (SOKENDAI), Mitaka, Tokyo 181-8588, Japan}

\author[0000-0001-9290-7846]{Ruobing Dong}
\affil{Department of Physics \& Astronomy, University of Victoria, Victoria, BC, V8P 5C2, Canada}

\author{Takayuki Muto}
\affil{Division of Liberal Arts, Kogakuin University, 1-24-2, Nishi-Shinjuku, Shinjuku-ku, Tokyo 163-8677, Japan}
\affil{Leiden Observatory, Leiden University, P.O. Box 9513, NL-2300 RA Leiden, The Netherlands}
\affil{Department of Earth and Planetary Sciences, Tokyo Institute of Technology, 2-12-1 Oh-okayama, Meguro-ku, Tokyo 152-8551, Japan}

\author[0000-0003-2300-2626]{Hauyu Baobab Liu}
 \affil{Department of Physics, National Sun Yat-Sen University, No. 70, Lien-Hai Road, Kaohsiung City 80424, Taiwan, R.O.C.}
 \affiliation{Center of Astronomy and Gravitation, National Taiwan Normal University, Taipei 116, R.O.C.}
 
\author[0000-0003-2887-6381]{Yuka Terada}
\affil{Institute of Astronomy and Astrophysics, Academia Sinica, 11F of Astronomy-Mathematics Building, AS/NTU No.1, Sec. 4,
Roosevelt Rd, Taipei 10617, Taiwan, R.O.C}
\affil{Department of Astrophysics, National Taiwan University, Taipei 10617, Taiwan, R.O.C.}

\begin{abstract}
An inner component misaligned from an outer component in a protoplanetary disk can result in the former casting shadows on the latter. 
We present a new instance of shadowing on the outer disk around a very low mass star, ZZ~Tau~IRS. Through the analysis of near-infrared (NIR) archival data at $\lambda=1.6$~$\mu$m acquired with the Wide Field Camera 3 on the Hubble Space Telescope, we identified brightness asymmetries in the top and bottom halves of the highly inclined outer disk, separated by a dark lane. The brighter sides in the top and bottom halves are on the opposite sides, which we attributed to shadows cast by a misaligned inner disk. Radiative transfer modeling of the system with a misaligned angle of 15~deg between the inner and outer disks well reproduced the observations. 
Additionally, we found an elevated brightness temperature of $^{12}$CO~(3-2) at $r\sim30$~au on the brighter side in NIR wavelengths in the top half disk, which can be explained by the shadowing effect too. 
While the origin of the misaligned inner disk remains unclear, future monitoring observations to search for temporal variations in brightness asymmetries will likely provide useful clues.
\end{abstract}

\keywords{Protoplanetary disks (1300); Planetary-disk interactions (2204); Exoplanet formation (492); M stars (985); Near infrared astronomy (1093); Hubble Space Telescope (756); stars: individual (ZZ Tau IRS)}

\section{Introduction} \label{sec:intro}

Planet formation takes place in protoplanetary disks. During their formation, gas giant planets perturb the host disk and produce substructures through planet-disk interactions \citep[e.g.][]{Paardekooper2023PPVII}. A variety of substructures, including rings/gaps, spirals, and crescents, have been predicted \citep[e.g.][]{Bae2023PPVII}, and observed in numerous disks, with rings/gaps being the most prevalent \citep[e.g.][]{Andrews2018DSHARP,vandermarel2023}. Additionally, some systems exhibit misaligned inner disks relative to outer disks \citep[e.g.,][]{Benisty2023PPVII}. It is noteworthy that the orbits of planets in our Solar System are also misaligned relative to the Sun's equator \citep[e.g.,][]{Kuiper1951SolarSystem}, which may be related to these misaligned planet-forming disks. 

Various mechanisms have been proposed to induce a misaligned inner disk. One such mechanism involves a misaligned stellar magnetic field relative to its disk, which can create a magnetically warped inner disk edge \citep[e.g.,][]{bouv99}. Anisotropic accretion flow of gas from envelopes is also expected to occur in early evolutionary stages such as class~0/I, resulting in subsequent formation of disks with different rotational axes \citep[e.g.,][]{Bate2018a, sakai2019, sai2020a}. Furthermore, a misaligned massive companion embedded in the disk can cause disk misalignment through companion-disk interactions \citep[e.g.,][]{Zhu2019a}.

Misaligned inner disks have been discerned through both direct and indirect methods. Infrared interferometric observations have directly constrained the geometries of a few misaligned inner disks \citep[e.g.,][]{Labdon2019CHARA,Bohn2022a}. Indirectly, shadows in the outer disk, cast by a misaligned inner disk obstructing starlight, have been detected at optical and near-infrared (NIR) wavelengths. While an inner disk with a small misalignment of a few to $\sim$10 degrees can cast a broad shadow in the outer disk \citep[e.g.,][]{Debes2017TWHya,beni2018,muro2020a,Zurlo2021BHB2007-1,Villenave2023Edge-on}, an inner disk misaligned by a few tens of degrees casts two narrow shadow lanes \citep[e.g.,][]{Krist2002GGTau,Marino2015HD142527,Pinilla2015J1604,Stolker2016a,Benisty2017HD100453,Casassus2018a,Kraus2020GWOri,Ginski2021SUAur}. Another observational approach capable of identifying misalignments in disks is gas kinematic observations at radio frequencies: the misaligned inner disk twists the gas velocity field, creating an `S'- or `Z'-shaped pattern at the boundary of red-shifted and blue-shifted components \citep{Rosenfeld2014Radialflow,Mayama2018,Bi2020GWOri}.

When the misaligned inner disk casts a shadow on the outer disk, local disk temperature in the shadow may or may not decrease significantly, depending on the cooling timescale. The temperature drops depend on the surface density of the outer disk, the extent of cooling during the shadow crossing time, and the efficiency of radiative diffusion \citep[e.g.,][]{Casassus2019Cooling,Arce-Tord2023DoAr44}. Analyses of shadowed disks provide an alternative method for estimating the surface density of disks.

In this paper, we report an occurrence of shadowing in the protoplanetary disk surrounding ZZ~Tau~IRS. ZZ~Tau~IRS, with a mass of 0.1 to 0.2~$M_\sun$ \citep{Andrews2013}, is classified as a very low-mass (VLM) star and is situated at an assumed distance of 130.7~pc\footnote{We do not use the GAIA distance of 105.7~pc because the astrometric solution is poor with a value of the renormalized unit weight error (RUWE) of 2.490 \citep[GAIA DR3;][]{gaia2016,gaia2022}.} \citep{akeson2019} in the Taurus star-forming region. It hosts a dust disk with a mass of about 24--50~$M_\earth$ \citep{Hashimoto2021zztauirs}, making it the most massive known disk among VLM stars \citep[e.g.,][]{ansdell2017a}. The outer ring at $r \sim 60$~au is also the largest, and it hosts a crescent with an azimuthal contrast of 1.4 at $\lambda = 0.9$~mm \citep{Hashimoto2021zztauirs}. Recent observations have revealed the growth of dust grains in the crescent to millimeter size \citep{Hashimoto2022ZZTauIRS}. ZZ~Tau~IRS is currently the only known VLM star with a ring and crescent configuration \citep[e.g.,][]{Pinilla2022review}. The innermost disk is likely nearly edge-on \citep{white2004,furlan2011} while the outer disk has an inclination of approximately 60~deg \citep{Hashimoto2021zztauirs}, suggesting a misalignment with the outer ring at an angle of approximately 30~deg. This fact implies the presence of shadows, which have not been previously reported.

\section{HST archive data and PSF subtraction} \label{sec:obs}

We utilized archival data of ZZ~Tau~IRS from observations conducted with the Wide Field Camera 3 (WFC3) on the Hubble Space Telescope (HST) as part of HST Proposal~14181 \citep[PI: T.~Megeath;][]{Pontoppidan2020HST,Habel2021HST} . The observations were executed on August 23, 2017, with an exposure time of 1,596.9 seconds. The broad-band filter F160W, centered on 1536.9~nm with a bandwidth of 268.3~nm, was employed for these observations. The data underwent calibration through the \verb#calwf3# data processing pipeline. The calibration process included basic calibration of the raw data such as subtracting the zeroth read, dark, and bias, correcting for detector nonlinearities, performing flat fielding, and removing cosmic rays. As the \verb#calwf3# pipeline does not correct for geometric distortion, \verb#AstroDrizzle#, part of the \verb#Drizzlepac# package, was utilized for distortion correction. The final product, suffixed with `\verb#_drz#,' was obtained from the Multimission Archive at STScI (MAST). The stellar Full Width at Half Maximum (FWHM), measured in field stars within the field of view (FoV), was found to be 0$\farcs$23. Finally, the absolute flux, expressed in Jy, was determined using the value of \verb#PHOTFNU# in the FITS header.

\begin{figure*}[htbp]
         \includegraphics[width=\linewidth]{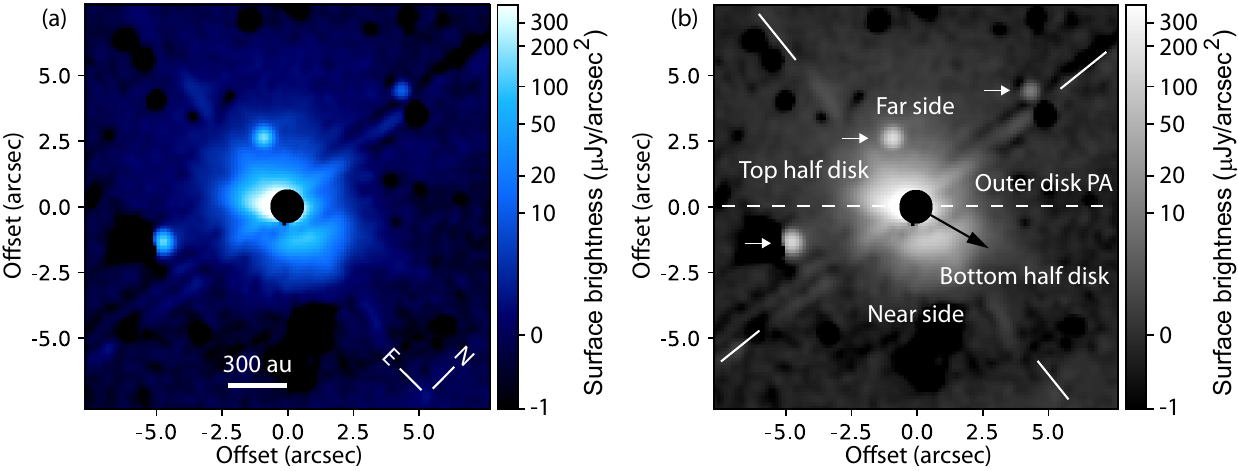} 
    \caption{
    \jhrev{The 1.6~$\mu$m NIR images of ZZ~Tau~IRS after PSF-subtraction.}
    The two panels show the same image with different annotations. The image is convolved by a Gaussian kernel with a size of 0\farcs64 in diameter and rotated by $-44.73$~deg to align the previously known outer disk PA horizontally, indicated by a horizontal dashed white line in panel~(b). The central region is masked due to high residual noise. Black spots represent background stars added in PSF subtraction for ZZ~Tau~IRS and ZZ~Tau, which is located at 35$\arcsec$ to the north. Three positive sources pointed out by white arrows in panel~(b) are also background stars because their movements between observations at 1999-Oct-25 \citep[Hubble Source Catalog;][]{Whitmore2016HSTcatalog} and 2017-Aug-23 (this study) are more than 1$\arcsec$ while it is approximately 0\farcs3 for ZZ~Tau~IRS. Four white bars at the corners indicate the direction of the diffraction pattern of the telescope spider. Black arrow represents the direction of the radial profiles at PA~$=$~265~deg shown in Figure~\ref{fig:radprof}.
    }
    \label{fig:zztauirs_image}
\end{figure*}

Prior to subtracting the Point Spread Function (PSF) of ZZ~Tau~IRS, the PSF of ZZ~Tau, situated 35$\arcsec$ to the north \jhrev{(refer to the HST image with a wider field-of-view in Figure~\ref{figA:field} in Appendix),
was subtracted using the Image Reduction and Analysis Facility \citep[IRAF;][]{Tody1986IRAF}.}
This step was necessary because the diffraction pattern of the telescope spider associated with ZZ~Tau contaminated the PSF of ZZ~Tau~IRS. The PSF used for subtracting ZZ~Tau is at the same filter F160W and was meticulously chosen from the WFC3 PSF database\footnote{https://www.stsci.edu/hst/instrumentation/wfc3/data-analysis/psf/psf-search} on the MAST Portal. We selected a PSF that exhibited sufficient brightness to induce a bright diffraction pattern at the position of ZZ~Tau~IRS. Additionally, to avoid subtractions on the disk of ZZ~Tau~IRS, the PSF used to subtract ZZ~Tau was chosen to have no visible stars within 3$\arcsec$ at the position of ZZ~Tau~IRS. U~Car observed with WFC3/F160W as part of HST Program 13335 \citep[PI: A.~Riess;][]{Riess2018HST1,Riess2018HST2} was chosen for PSF subtraction of ZZ~Tau by simply scaling the PSF of U~Car. For the PSF subtraction of ZZ~Tau~IRS, we selected a PSF star, Gaia DR3 147869474824734208, in the same image as ZZ Tau IRS, with no visible stars within 3$\arcsec$. The PSF of ZZ Tau IRS was then subtracted by scaling the PSF star. The PSF-subtracted image is shown in Figure~\ref{fig:zztauirs_image}. 
\jhrev{
While the spatial resolution is 0\farcs23, we convolved the image with a larger Gaussian kernel of 0\farcs64 in diameter to mitigate residuals from the PSF subtraction process, specifically addressing remnants of the diffraction pattern from the telescope spider.
} 
As several stars are contaminated in the PSF stars, there are black spots beyond 3$\arcsec$ from ZZ~Tau~IRS in the PSF-subtracted image.

\section{Results} \label{sec:results}

Extended structures out to approximately $r=2\farcs5$ (300~au) have been identified around ZZ~Tau~IRS in the scattered light image in Figure~\ref{fig:zztauirs_image}(a). The top and bottom half disks are separated by a dark lane annotated in Figure~\ref{fig:zztauirs_image}(b), representing the disk midplane of the disk. This result is consistent with the fact that the outer disk at $r\sim60$~au in ALMA observations is known to be inclined by approximately 60~deg \citep{Hashimoto2021zztauirs}. 
The dark lane appears to exhibit a slight diagonal orientation along the south-east/north-west axis due to the absence of a portion of the dark lane. This effect is also evident in our modeling in \S \ref{sec:modeling}.

\begin{figure}[htbp]
         \includegraphics[width=\linewidth]{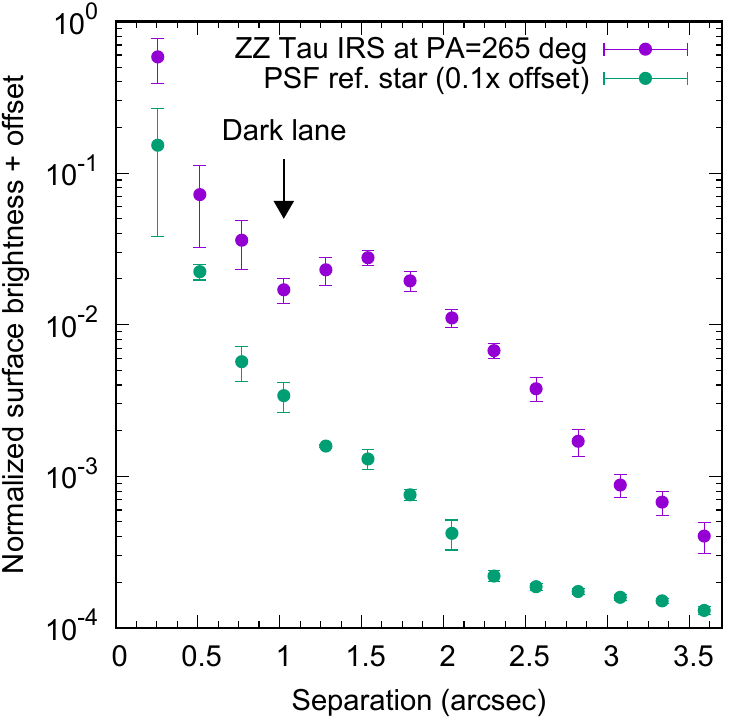} 
    \caption{
      Radial profiles along PA~$=$~265~deg, as indicated by the black arrow in Figure \ref{fig:zztauirs_image}(b), for the PSF-subtracted ZZ~Tau~IRS and the PSF reference star Gaia DR3 147869474824734208. A dip, representing the dark lane, is clearly observed at approximately $r=1$~arcsec for ZZ~Tau~IRS, 
      \jhrev{while the intensity profile of the PSF reference star exhibits a monotonic decrease with radial distance.}
    }\label{fig:radprof}
\end{figure}

We assessed the existence of the dark lane by examining the radial profiles. Figure~\ref{fig:radprof} displays the radial profiles along PA~$=$~265 deg, which is the direction indicated by the black arrow in Figure~\ref{fig:zztauirs_image}(b), for the PSF-subtracted ZZ~Tau~IRS and the PSF reference star Gaia DR3 147869474824734208.  Since the dark lane along the disk minor axis at PA~$=$~225 deg is closer to the mask edge, we opted to plot the radial profiles at PA~$=$~265 deg instead. We distinctly observed a dip at approximately $r=1$~arcsec in the radial profile of ZZ~Tau~IRS. As the PSF reference star exhibits monotonic decrease, the dip in the radial profile of ZZ~Tau~IRS is not attributed to artifacts created by the PSF subtraction processes outlined in \S~\ref{sec:obs}. Therefore, we consider that the dark lane is a real feature in the disk.

The top and bottom half disks exhibit brightness asymmetries, however on the opposite sides. The southeast side is brighter in the top half, while the northwest side is brighter in the bottom half. We attribute these brightness asymmetries to shadows cast by the misaligned inner disk on the outer disk. Such a brightness asymmetry caused by shadows has been reported in disk modeling before, for example, by \citet{Facchini2018Warp} (refer to their Figure~9m). 

\begin{figure}[htbp]
         \includegraphics[width=\linewidth]{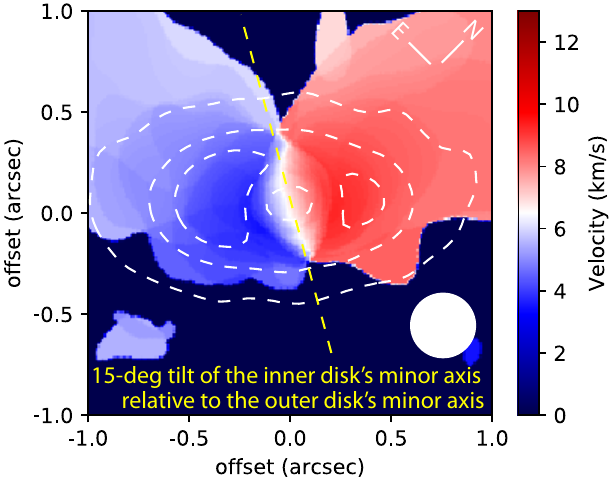} 
    \caption{
     The $^{12}$CO~(3-2) moment~1~map adapted from \citet{Hashimoto2021zztauirs} with the beam size of 329~$\times$~327~mas indicated by the ellipse in right bottom corner. The image is rotated with $-44.73$~deg to align the disk PA horizontally. The dashed yellow line represents a direction 15~deg away from the outer disk's minor axis. The white dashed contours represent the dust continuum at $\lambda~=~0.9$~mm at 10$\sigma$, 30$\sigma$, and 50$\sigma$ (1$\sigma=268$~$\mu$Jy/beam) taken from \citet{Hashimoto2021zztauirs}.
    }\label{fig:co_m1}
\end{figure}

To explore additional observational evidence of the presence of a misaligned inner disk, we revisited ALMA $^{12}$CO~(3-2) observations reported by \citet{Hashimoto2021zztauirs}. 
In cases where the inner disk is misaligned with the outer disk, the boundary between red-shifted and blue-shifted components in the inner disk --- i.e., the minor axis of the inner disk --- is expected to be tilted relative to the outer disk's minor axis \citep[e.g.,][]{Rosenfeld2014Radialflow}, unless the minor axes of the inner and outer disks are aligned.
Figure~\ref{fig:co_m1} shows the $^{12}$CO moment~1~map. We found an approximate 15-deg tilt of the minor axis in the inner disk relative to the outer disk's minor axis. While acknowledging that the 15-deg tile of the inner disk's minor axis in Figure~\ref{fig:co_m1} may be affected by the beam size, we are confident in asserting the presence of a misaligned inner disk around ZZ~Tau~IRS.

We did not find a cavity/gap structure in the HST image in Figure~\ref{fig:zztauirs_image}, while a dust ring structure at $r=58$~au (0$\farcs$45) was previously reported in (sub)mm wavelengths \citep[][refer to Figure~\ref{figA:alma} in Appendix]{Hashimoto2021zztauirs}. The central region within approximately $r=80$~au (0$\farcs$6) in Figure~\ref{fig:zztauirs_image} is masked due to large residual noises. A potential cavity/gap in the optical/NIR wavelengths comparable to the mm-ring may be present. In addition, the counterpart of the (sub)mm emission cavity at the NIR wavelength may have a small radius \citep{dong12cavity}, making it harder to detect.

Additionally, no narrow shadows are apparent in the outer disk in Figure~\ref{fig:zztauirs_image}. 
\citet{white2004} and \citep{furlan2011} have suggested the presence of a nearly edge-on innermost disk, which is expected to cast two narrow shadows on the outer disk
\citep[refer to Figure~26 in][]{Whitney2013MCRT}. 
The identification of such narrow shadows in the present HST observations may be challenging due to the low angular resolution of 0\farcs23. To more effectively search for the cavity/gap structure and the two narrow shadows, higher spatial and contrast observations are necessary.

\section{Radiative transfer modeling} \label{sec:modeling}

As introduced in \S~\ref{sec:intro}, the brightness asymmetry in the outer disk can be attributed to shadows cast by a misaligned inner disk. Given that nearly half of the top and bottom half outer disks are shadowed in Figure~\ref{fig:zztauirs_image}, the misalignment is likely small \citep[e.g.,][]{Debes2017TWHya,beni2018,muro2020a}. In this section, we assess the aforementioned hypothesis by employing radiative transfer modeling through the Monte Carlo radiative transfer (MCRT) code (\verb#HOCHUNK3D#; \citealp{Whitney2013MCRT}).

\begin{figure}[htbp!]
    \includegraphics[width=\linewidth]{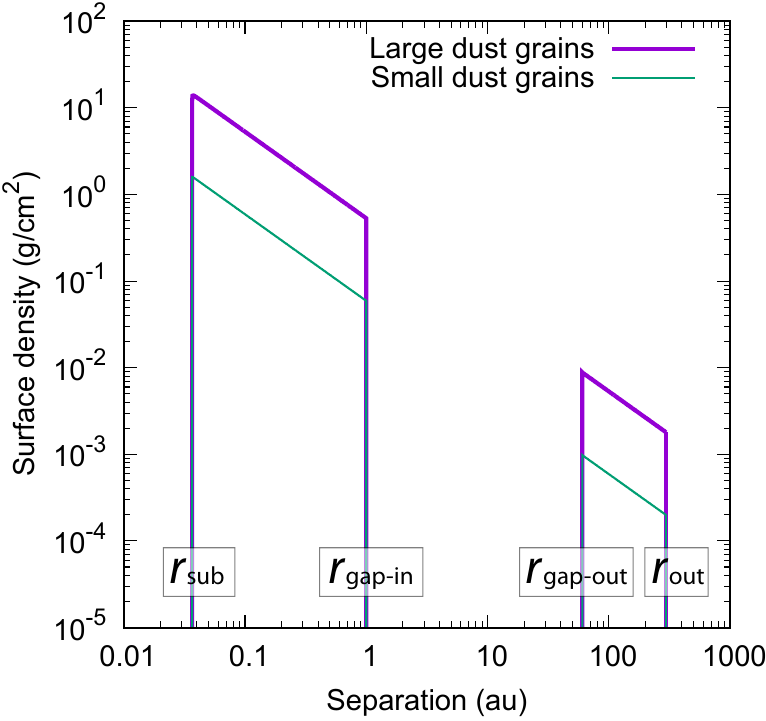} 
    \caption{
    The surface density of large and small dust grains in our model. 
    }\label{fig:sigma}
\end{figure}

In our modeling, we excluded the edge-on innermost disk proposed by \citet{white2004,furlan2011} because the HST image in Figure~\ref{fig:zztauirs_image} did not distinctly reveal the two expected narrow shadows (\S~\ref{sec:results}). We considered two disks: the misaligned inner disk casting shadows and the outer disk shadowed over extended areas.

\subsection{Setup and results}\label{subsec:setup}

The MCRT code follows a two-layer disk model featuring small dust grains (up to micron size) well-mixed with the gas and large dust grains (up to millimeter size) in the disk midplane \citep[e.g.,][]{DAlessio2006Disk}. The disk structure and dust properties are described in \citet{hashimoto+15}. In summary, we utilize the standard interstellar-medium dust model for the small dust grains (\citealt{Kim1994SmallDust}; a composite of silicates and graphites with a size distribution of $n(s) \propto s^{-3.5}$ from $s_{\rm min} = 0.0025$~$\mu$m to $s_{\rm max} = 0.2$~$\mu$m) and dust Model~2 in \citet{Wood2002LargeDust} for the large dust grains (a composite of carbons and silicates with a size distribution of $n(s) \propto s^{-3.5}$ from $s_{\rm min} = 0.01$~$\mu$m to $s_{\rm max} = 1000$~$\mu$m). 

\begin{figure}[htbp!]
    \includegraphics[width=\linewidth]{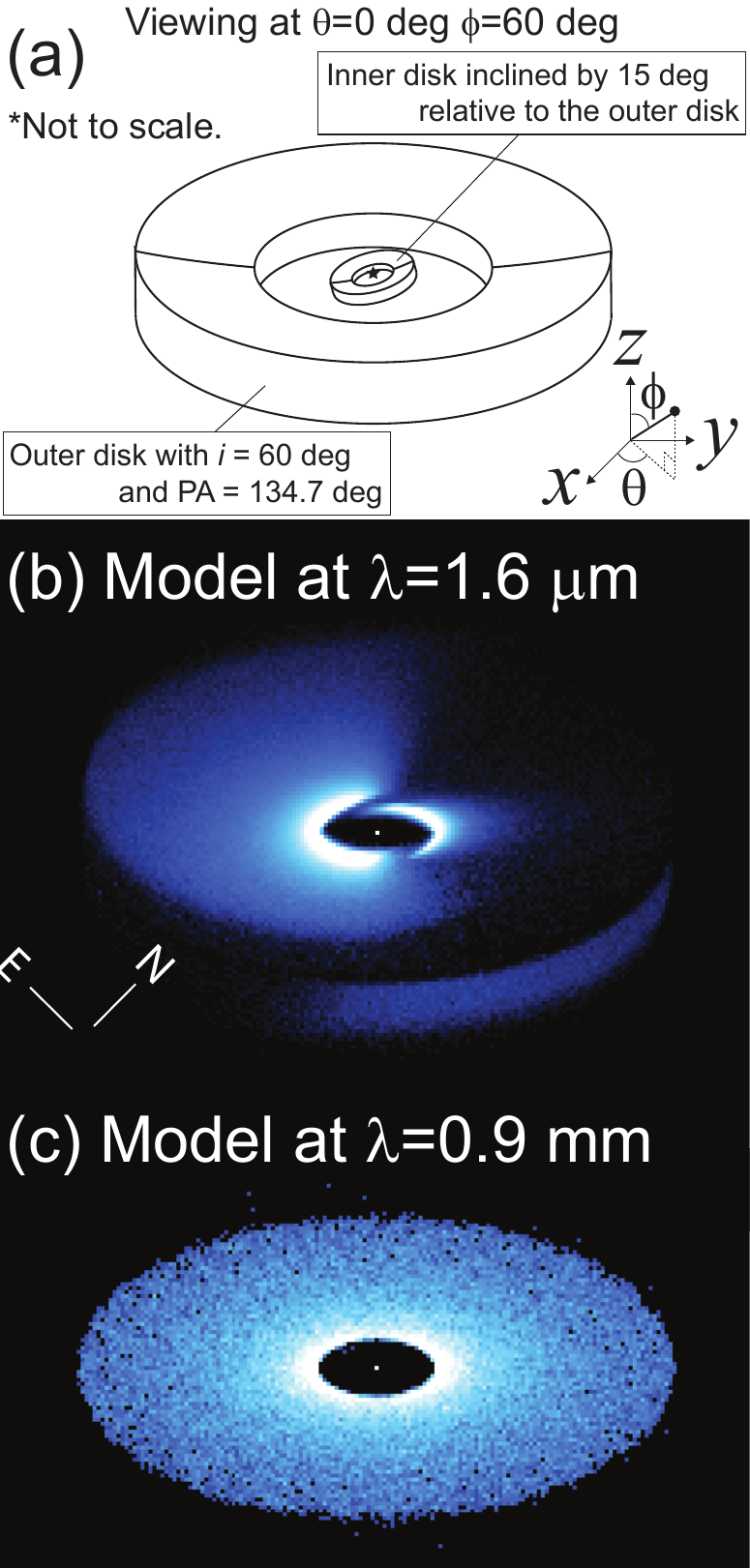} 
    \caption{
    (a) Schematic view of the geometry in the inner and outer disks in our modeling. The inner and outer disks are separated by the gap at $r=1$ to 60~au. The inner disk is inclined by 15~deg relative to the outer disk. The whole system is viewed at $\theta=0$~deg and $\phi=60$~deg. Refer to text in \S~\ref{subsec:setup} for details.
    (b and c) Synthesised scattered light image at $\lambda=1.6$~$\mu$m and dust continuum image at $\lambda=0.9$~mm at raw resolution in our modeling. 
    }\label{fig:mcrt}
\end{figure}

\begin{figure*}[htbp!]
\centering
    \includegraphics[width=0.8\linewidth]{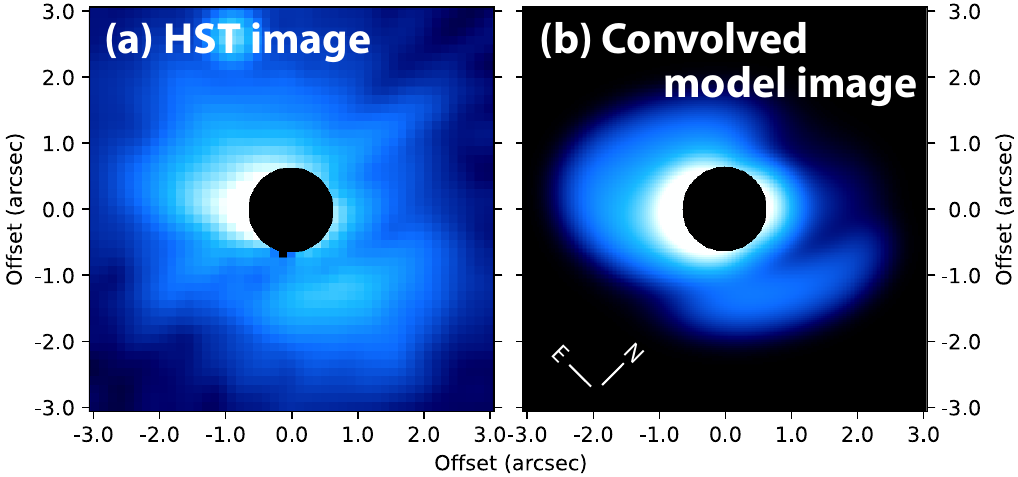} 
    \caption{
    Comparison between the HST image and the convolved model image. (a) The HST image, as shown in Figure \ref{fig:zztauirs_image}(a). (b) The model image from Figure \ref{fig:mcrt}(b) convolved with a Gaussian kernel with a diameter of 0\farcs64, same as the Gaussian kernel applied in panel (a).
    }\label{fig:comparison}
\end{figure*}

The radial surface densities for both large and small grains shown in Figure~\ref{fig:sigma} were assumed to follow the same power-law profiles:
\begin{eqnarray*}
\Sigma(r) &=& C \cdot \Sigma_{0} (r/r_{0})^{-1}, \\
C &=& \left \{
\begin{array}{llllllll}
0 &{\rm for}&                 & & r  &\leq& r_{\rm sub}\\
1 &{\rm for}& r_{\rm sub}     &<& r  &\leq& r_{\rm gap\mathchar`-in}\\
0 &{\rm for}& r_{\rm gap\mathchar`-in}  &<& r  &\leq& r_{\rm gap\mathchar`-out}\\
1 &{\rm for}& r_{\rm gap\mathchar`-out} &<& r  &\leq& r_{\rm out},
\end{array}
\right. 
\end{eqnarray*} 
where $r_{0}$ is a radial scaling factor, and $\Sigma_{0}$ is the surface density at $r_0$ set by the total (gas~$+$~dust) disk mass ($M_{\rm disk}$) assuming a gas-to-dust mass ratio of 100.
The dust sublimation radius $r_{\rm sub}$ is at approximately 0.04~au and is determined as the radius at which the temperature reaches the sublimation temperature $T_{\rm sub} \sim 1600$~K \citep{Dullemond2001}. 

We introduce a gap that separates the inner and outer disks. To simplify the model, we assume this gap to be devoid of material. In practice, the residual gas within the gap may form a warp connecting the inner and outer disks, resulting in the observed CO emission pattern (Figure~\ref{fig:co_m1}). Regarding the inner disk, given its poorly constrained size, we assume $r_{\rm gap\mathchar`-in}=1$~au. Our experiments have shown that the shadow in the outer disk is not affected by the inner disk size. For the outer disk, we adopt an inner radius of $r_{\rm gap\mathchar`-out}=60$~au based on ALMA dust observations \citep{Hashimoto2021zztauirs}. The outer edge of the outer disk is assumed to be at $r_{\rm out}=300$~au, as indicated by Figure~\ref{fig:zztauirs_image}. Figure~\ref{fig:sigma} illustrates the surface densities of large and small dust grains in our model. We set $M_{\rm disk}$ to 0.01~$M_\sun$ based on previous ALMA (sub)mm observation \citep{Hashimoto2021zztauirs}. The mass fraction of large dust grains in the total dust mass is set to 0.9.

The scale heights ($h$) of both large and small grains were assumed to change radially following $h \propto r^{p}$. We assumed $p=1.15$ as it yielded a favorable fit to the observations. At $r=100$~au, we set $h$ to be 1~au for the large dust grain disk and 7~au for the small dust grain disk.

To illustrate the geometry of misalignment, we introduce a Cartesian coordinate system, as depicted in the inset of Figure~\ref{fig:mcrt}(a). Initially, the midplane of both the inner and outer disks are positioned on the xy-plane. Next, the rotational axis of the inner disk, originally pointing in the $z$-direction, is tilted within the $yz$-plane by 5 to 30~deg towards the $-y$-direction. After several iterations, we set the tilt angle to 15~deg to achieve a good agreement with HST observations. 
Finally, we observe the system at $\theta=0$~deg and $\phi=60$~deg (i.e., the direction of the line of sight), representing a disk inclination of 60~deg \citep{Hashimoto2021zztauirs}, as shown in the schematic view in Figure~\ref{fig:mcrt}(a).

Figure~\ref{fig:mcrt}(b) presents a synthesized scattered light image at $\lambda=1.6$~$\mu$m of our model. The top and bottom half disks show brightness asymmetries on the opposite sides, reproducing the morphology of the disk around ZZ~Tau~IRS in Figure~\ref{fig:zztauirs_image}. 
While the outer disk in our model is set to align perfectly horizontally, the dark lane representing the disk midplane appears to exhibit a slight diagonal orientation along the southeast/northwest axis due to the effect of shadows. This is consistent with the observed image in Figure~\ref{fig:zztauirs_image}. Figure~\ref{fig:comparison} presents a direct comparison between the HST image and the model image convolved with a Gaussian kernel with a diameter of 0\farcs64, demonstrating a good agreement between the two.

\begin{figure*}[htbp]
         \includegraphics[width=\linewidth]{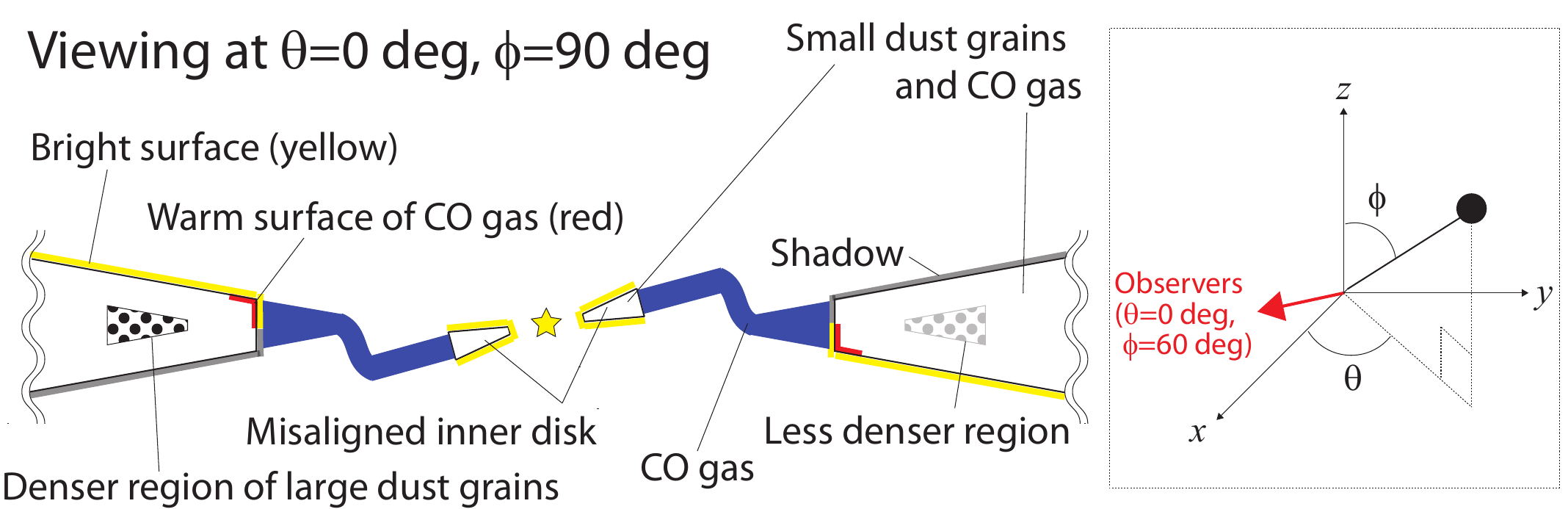} 
    \caption{
    Schematics of the ZZ~Tau~IRS system in an edge-on view, i.e., viewed at $\theta=0$~deg and $\phi=90$~deg. Earth-bound observers view the disks at $\theta=0$~deg and $\phi=60$~deg. In addition to the misaligned inner disk and the shadowed outer disk, the outer disk of large dust grains (dots), warmer CO gas (red), and CO gas in the gap (blue) are illustrated. Refer to the text in \S~\ref{subsec:setup} and \S~\ref{subsec:alma} for further details.
    }
    \label{fig:schematics}
\end{figure*}

\subsection{Caveat}

As mentioned at the beginning of this section, we excluded the edge-on inner disk proposed by \citet{white2004,furlan2011} from our modeling. The Spectral Energy Distribution (SED) of ZZ~Tau~IRS clearly reveals deficits in flux density at optical/NIR wavelengths \citep{furlan2011}, possibly caused by the blocking of stellar light by the edge-on inner disk. In contrast, since the inclination of the misaligned inner disk in our modeling is approximately 60~deg, the central star is not significantly obscured by the misaligned inner disk \citep[see Figure~2 in][]{Whitney2013MCRT}. Although our modeling effectively replicates the observed asymmetric features in the outer disk captured by HST, it does not reproduce the SED. A comprehensive exploration of the parameter space is necessary to assess the feasibility of adjusting the properties of the inner and outer disks in our model (e.g., viewing geometry, scale height, and mass) to reconcile both the HST observations and the SED. Alternatively, it is plausible that an additional disk component, such as an innermost disk inside our shadow-casting disk, is required to adequately reproduce the SED. These considerations extend beyond the scope of this paper.

\section{Discussion} \label{sec:discuss}

\subsection{Comparison with ALMA (sub)mm results}\label{subsec:alma}

ZZ~Tau~IRS was observed with ALMA at $\lambda=870$~$\mu$m \citep{Hashimoto2021zztauirs}. We have presented a comparison between the NIR image (Figure~\ref{fig:zztauirs_image}) and ALMA images showing the $^{12}$CO brightness temperature and the dust continuum in Figure~\ref{figA:alma} in the Appendix. Both ALMA images relate to temperature, exhibit greater brightness on the southeast side. The $^{12}$CO is generally optically thick, effectively tracing the surface of the upper disk. The elevated surface temperature on the southeast side of the top disk in Figure~\ref{figA:alma}(a) can be attributed to direct irradiation, a consequence of the shadowing inner disk.

In the case of dust continuum, significant contributions to (sub)millimeter emissions come from large dust grains situated in the disk midplane \citep[e.g.,][]{Andrews2020ARAA}. The disk composed of these large dust grains is geometrically thin, allowing it to be considered vertically isothermal. Despite the presence of shadows, both the southeast and northwest sides receive an equal amount of irradiation, resulting in comparable disk temperature on both sides. This is different from the situation in which narrow lane shadows result in a local temperature drop relative to the rest of the disk \citep[e.g.,][]{Casassus2019Cooling,Arce-Tord2023DoAr44}. Consequently, the observed brightness asymmetry in the dust continuum cannot be attributed to shadows. Indeed our modeling in dust continuum at $\lambda=0.9$~mm in Figure~\ref{fig:mcrt}(c) exhibits symmetric brightness distributions. Instead, it may be linked to asymmetric distributions of properties and/or surface density of dust grains.

Figure~\ref{fig:schematics} presents schematics of the ZZ~Tau~IRS system. In addition to featuring the misaligned inner disk and the shadowed outer disk, the diagram depicts the outer disk composed of large dust grains, traced by (sub)millimeter emissions. It also shows the presence of warmer CO gas in proximity to the central star, along with CO gas within the gap that might form a warp connecting the inner and outer disks.

\subsection{Origin of misalignment}

Several mechanisms have been proposed to elucidate disk misalignments, such as the stellar magnetic field, anisotropic accretion flow, and the influence of an inclined massive companion (refer to \S~\ref{sec:intro}). The variation in shadows caused by the precession of the inner disk periodically obscuring the central star could serve as a diagnostic tool for identifying the origins. However, the presence of such variability in the outer disk around ZZ~Tau~IRS has not been explored, as only one epoch of HTS/WFC3 NIR observation has been carried out\footnote{Except HTS/WFC3 NIR observations, ZZ~Tau~IRS was observed with HST/WFPC2 at visible wavelengths in 1999-Oct-25 (program ID of 8216). The comparison of ZZ~Tau~IRS disk at different wavelengths will be present in the future work.}. In this subsection, we will explore how to differentiate the origin of misalignment through the brightness variability of the outer disk.

\emph{Stellar magnetic field ---} 
The inner disk undergoes warping due to a misaligned stellar magnetic field, and the timescale of variability may align with the stellar rotational period. However, the rotational period of ZZ~Tau~IRS has not been reported. Assuming that ZZ~Tau~IRS shares a similar rotational period with other T-Tauri stars \citep{Bouvier1989StarRot}, it could be approximately 1-10 days.

\emph{Anisotropic accretion flow ---} 
The direction of angular momentum in different disks is anticipated to be stable on short timescales, as precession of disks caused by their mutual gravitational interaction is expected to be slow due to the large separation of the inner and outer disks in the current model. 

\emph{A massive companion on an orbit misaligned from the disk ---} 
Such a companion has the potential to create a gap and induce misalignment and precession of the inner disk \citep[e.g.,][]{Zhu2019a}. According to equation~(28) in \citet{Zhu2019a}, the precession timescale for $q=M_{\rm p}/M_*=0.05$ is approximately 10$^2$ orbits of the planet. Even if the 10~$M_{\rm Jup}$ planet's orbit is as close as 0.05~au to the dust sublimation radius (\S~\ref{sec:modeling}), the precession timescale for the inner disk around ZZ~Tau~IRS is approximately 10$^3$ days. 

According to the core (pebble) accretion scenario, very low-mass (VLM) stars, typically defined as having a mass of $\lesssim0.2 M_\sun$, including ZZ~Tau~IRS, are expected to be capable of forming only low-mass planets, typically up to a few Earth masses \citep{Payne2007a,Ormel2017a,Schoonenberg2019,liu2019a,Liu2020a,miguel2020a,burn2021}. Meanwhile, a recent study by \citet{Pan2023VLM} suggests that even around VLM stars, giant planets are still likely to form. This process involves the initial formation of multiple low-mass protoplanets, followed by their growth through a combination of pebble accretion and planet-planet collisions within disks.

The origin of misalignment of the inner disk around ZZ~Tau~IRS remains uncertain. Meanwhile, monitoring observations for brightness asymmetries of the disk could provide valuable insights into its origin, as discussed earlier. Intriguingly, the discovery of long-term variability with a timescale of approximately 10$^3$~days in the future could suggest the presence of a massive planet, although the situation is degenerated, as a brown dwarf at larger separation may lead to the same precession timescale of the inner disk as a planet at smaller separation. Such a finding would pose a challenge to existing core accretion models if the perturber is a massive planet.

\section{Summary}

We analyzed HST archival NIR ($\lambda=1.6$~$\mu$m) observations of ZZ~Tau~IRS. The scattered light image clearly shows that the disk is separated by a dark lane, typical for disks at relatively high inclinations. Additionally, we identified brightness asymmetries in the top and bottom halves of the disk, and their brighter sides on the opposition sides. We attributed the brightness asymmetries to shadows cast by a misaligned inner disk. Furthermore, we presented a radiative transfer model to explain the HST observations with an inner disk inclined by 15 degrees relative to the outer disk.

We revisited $^{12}$CO brightness temperature and the dust continuum at $\lambda=870$~$\mu$m observed by ALMA. Both images revealed a brighter side in the southeast region of the disk, similar to the pattern observed in the near-infrared scattered light image. The asymmetry of $^{12}$CO brightness temperature can be attributed to shadows, while the asymmetry in the dust continuum may be linked to asymmetry in the spatial distribution of dust surface density and/or dust properties. 

We discussed potential origins for the misaligned inner disk, including the stellar magnetic field, anisotropic accretion flow, and an inclined massive companion. While the specific cause remains open, future monitoring observations for variability of brightness asymmetries hold the promise of offering valuable insights.

\begin{acknowledgments}

The authors thank the anonymous referee for a timely and constructive report.
The authors thank Yifan~Zhou for useful comments on data reduction. 
This study was supported by JSPS KAKENHI Grant Numbers 21H00059, 22H01274, 23K03463.
H.B.L. is supported by the National Science and Technology Council (NSTC) of Taiwan (Grant Nos. 111-2112-M-110-022-MY3).
Some/all of the data presented in this article were obtained from the Mikulski Archive for Space Telescopes (MAST) at the Space Telescope Science Institute. The specific observations analyzed can be accessed via \dataset[DOI: 10.17909/etpa-dx81]{https://doi.org/10.17909/etpa-dx81}.
This paper makes use of the following ALMA data: ADS/JAO.ALMA\#2016.1.01511.S. ALMA is a partnership of ESO (representing its member states), NSF (USA), and NINS (Japan), together with NRC (Canada), NSC (Taiwan), ASIAA (Taiwan), and KASI (Republic of Korea), in cooperation with the Republic of Chile. The Joint ALMA Observatory is operated by ESO, AUI/NRAO, and NAOJ.
IRAF is distributed by the National Optical Astronomy Observatories, which are operated by the Association of Universities for Research in Astronomy, Inc., under cooperative agreement with the National Science Foundation.

\end{acknowledgments}

\software{
          astropy \citep{Astropy2013,Astropy2018,Astropy2022},  
          Numpy \citep{VanDerWalt2011}, 
          IRAF \citep{Tody1986IRAF},
          Matplotlib \citep{Hunter2007Matplotlib}
          }

\facilities{HST, ALMA}

\appendix

\section{HST image with wide field-of-view}\label{secA:field}

\begin{figure}[htbp]
         \includegraphics[width=\linewidth]{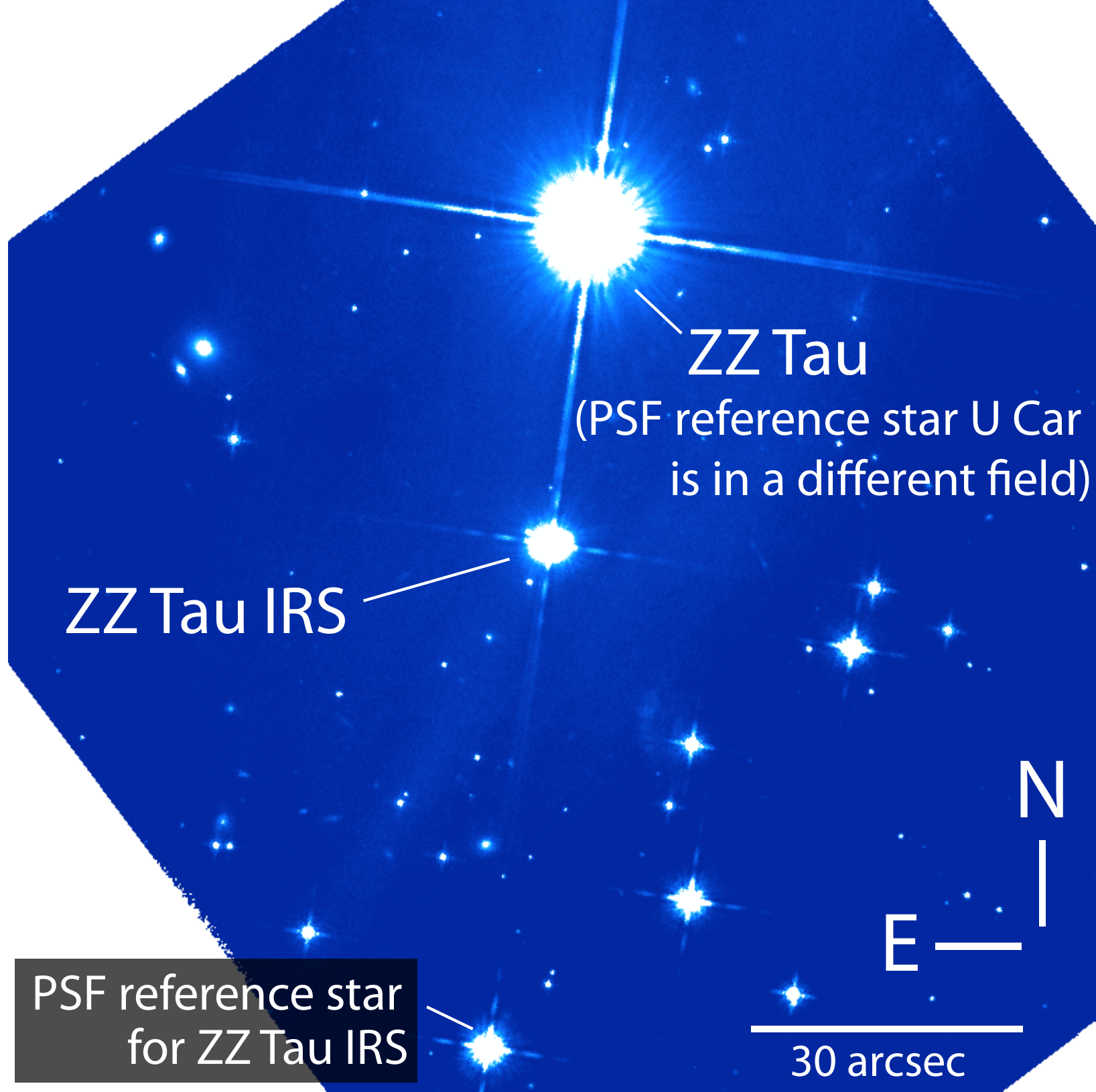} 
    \caption{
    The HST image with a wide field-of-view, encompassing ZZ~Tau, ZZ~Tau~IRS, and the PSF reference star Gaia DR3~147869474824734208 within the same field for ZZ~Tau~IRS. The PSF reference star for ZZ~Tau, U Car, is in a separate field.
    }
    \label{figA:field}
\end{figure}

We provide the HST image with a wide field-of-view \jhrev{in Figure~\ref{figA:field}} for readers to easily discern the positions of ZZ~Tau, ZZ~Tau~IRS, and the PSF reference star Gaia DR3 147869474824734208 for ZZ~Tau~IRS, \jhrev{in the PSF subtraction process in \S~\ref{sec:obs}}. The PSF reference star U~Car for ZZ~Tau is in a separate field.

\section{ALMA results of ZZ Tau IRS}\label{secA:alma}

\begin{figure*}[htbp]
         \includegraphics[width=\linewidth]{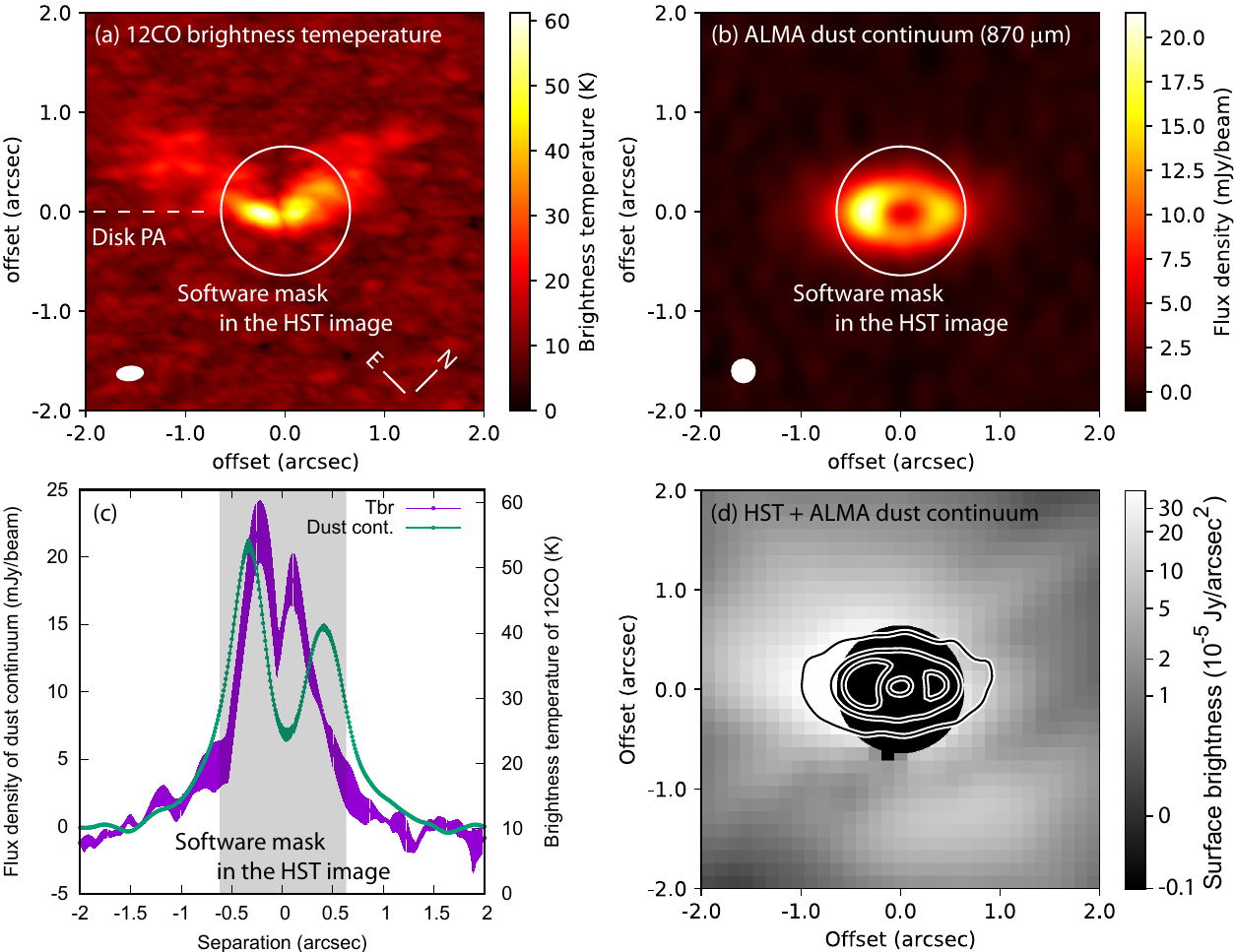} 
    \caption{
    ALMA images of dust continuum (a) and brightness temperature of $^{12}$CO~3-2 (b) for ZZ~Tau~IRS adapted from \citep{Hashimoto2021zztauirs}. These images are rotated with $-44.73$~deg to align the disk PA horizontally. (c) The radial profiles along disk PA in panels (a) and (b). (d) The contours representing the dust continuum at 10$\sigma$, 30$\sigma$, and 50$\sigma$ (1$\sigma=268$~$\mu$Jy/beam) are superposed on the HST image in Figure~\ref{fig:zztauirs_image}.
    }
    \label{figA:alma}
\end{figure*}

We revisited ALMA images of the $^{12}$CO brightness temperature and the dust continuum at $\lambda=870$~$\mu$m in Figure~\ref{figA:alma} \citep{Hashimoto2021zztauirs}.

\bibliography{sample63}{}
\bibliographystyle{aasjournal}

\end{document}